\documentclass[showpacs,amsmath,amssymb,10pt]{article}
\usepackage{graphicx,color}
\usepackage{amsmath}
\usepackage{amssymb}
\usepackage{amsmath,amsfonts,latexsym,graphicx,amssymb}
\usepackage{amsfonts}
\usepackage{amssymb}
\usepackage{mathrsfs}
\usepackage{amsthm}
\usepackage[utf8]{inputenc}
\usepackage{graphicx}
\usepackage{amsmath}
\newcommand{\ot}{{\,\otimes\,}}
\newcommand{{\Cd}}{{\mathbb{C}^d}}
\newcommand{{\Rn}}{{\mathbb{R}^n}}

\def\oper{{\mathchoice{\rm 1\mskip-4mu l}{\rm 1\mskip-4mu l}%
{\rm 1\mskip-4.5mu l}{\rm 1\mskip-5mu l}}}
\def\<{\langle}
\def\>{\rangle}
\newtheorem{thm}{Theorem}
\newtheorem{cor}{Corollary}

\newtheorem{definition}{Definition}

\newtheorem{pro}{Proposition}
\newtheorem{remark}{Remark}
\date{}

\begin{document}
\title{\textbf{Quantum-correlation breaking channels,  quantum conditional probability and Perron-Frobenius theory}}
\author{Dariusz Chru\'sci\'nski \\
Institute of Physics, Faculty of Physics, Astronomy and Informatics \\
 Nicolaus Copernicus University \\
Grudziadzka 5, 87--100 Toru\'n, Poland}

\maketitle

\begin{abstract}
Using the quantum analog of conditional probability and classical Bayes theorem we discuss some aspects of particular entanglement breaking channels: quantum-classical and classical-classical channels.  Applying the quantum analog of Perron-Frobenius theorem we generalize the recent result of Korbicz et. al. \cite{KHH} on full and spectrum broadcasting from quantum-classical channels to arbitrary quantum channels.
\end{abstract}




\section{Introduction}

Quantum channels provide the basic ingredients of modern Quantum Information Theory \cite{QIT}.
A channel with an input state $\mathcal{H}_A$ and an output state $\mathcal{H}_B$ is a linear completely positive trace preserving map $\Lambda_{B|A} : \mathfrak{T}(\mathcal{H}_A) \rightarrow \mathfrak{T}(\mathcal{H}_B)$, where $\mathfrak{T}(\mathcal{H})$ denotes a convex set of trace class operators in $\mathcal{H}$ \cite{Paulsen} (an operator $X \in \mathfrak{T}(\mathcal{H})$ iff $||X||_1 = {\rm Tr}\sqrt{X^\dagger X} < \infty$).  Note that in finite dimensional case considered in this Letter $\mathfrak{T}(\mathcal{H})$ coincides with $\mathfrak{B}(\mathcal{H}) = M_n(\mathbb{C})$ and $n = {\rm dim}\, \mathcal{H}$. Quantum states are represented by density operators belonging to  $\mathfrak{S}(\mathcal{H}) = \{ \rho \in \mathfrak{T}(\mathcal{H})\ ; \ \rho \geq 0 \ , \ {\rm Tr}\, \rho=1\}$.

An interesting subclass of channels is provided by so called entanglement breaking channels \cite{EB,Holevo} (see \cite{Holevo-infty} for the infinite dimensional case). A channel $\Lambda_{B|A}$ is entanglement breaking (EB) if for arbitrary Hilbert space $\mathcal{H}_R$ and arbitrary state $\rho_{RA} \in \mathfrak{S}(\mathcal{H}_R \ot \mathcal{H}_A)$ the output $(\oper_R \ot \Lambda_{B|A})\rho_{RA}$ defines a separable state in $\mathcal{H}_R \ot \mathcal{H}_B$. As usual $\oper_R$ is an identity map in $\mathfrak{T}(\mathcal{H}_R)$. This class was generalized to so called partially entanglement breaking channels \cite{PART}: a channel $\Lambda_{B|A}$ is $r$-PEB ($r$ partially entanglement breaking), if for any state $\rho_{RA}$ one has ${\rm SN}[(\oper_R \ot \Lambda_{B|A})\rho_{RA}] \leq r$, where ${\rm SN}[\rho_{RB}]$ denotes a Schmidt number of the density operator $\rho_{RB}$ \cite{SN}. It is clear that $1$-PEB are simply EB channels. In a recent paper \cite{KHH}
authors provided an interesting refinement of the characterization of EB channels
 to more general quantum correlations and connected it to measurement maps, quantum
state broadcasting, and finite Markov chains. By a quantum-correlation breaking channel we mean a quantum channel $\Lambda_{B|A}$ such that for any bi-partite input state $\rho_{RA}$ the output state $(\oper_R \ot \Lambda_{B|A})\rho_{RA}$ is less correlated than $\rho_{RA}$.

In this Letter we show that a natural framework to discuss this characterization is
the notion of  {\em quantum conditional probability} \cite{Manolo} (see also \cite{Hirota}) [or {\em conditional quantum states} \cite{Leifer,Leifer-2}]. Slightly different approach was also proposed in Ref.  \cite{Cerf}. Moreover, applying the quantum analog of Perron-Frobenius theorem we generalize  the analysis on full and spectrum broadcasting from quantum-classical channels \cite{KHH} to arbitrary quantum channels.

\section{Quantum conditional probability}

As is well known conditional probability plays important role in classical probability theory. Having two random variables $A$ and $B$ one introduces conditional probability (for $B$ given $A$) $p_{i|j} = P(B=i|A=j)$ which satisfies:
\begin{equation}\label{C1}
p_{i|j} \geq 0 \ ; \ \ \sum_i p_{i|j}=1\ .
\end{equation}
Knowing probability distribution $p^A_j$ for $A$ one has for the joint probability $p_{ij} = P(B=i,A=j)$
\begin{equation}\label{C2}
    p_{ij} = p_{i|j}\, p^A_j\ ,
\end{equation}
and hence the marginal probability for $B$ reads
\begin{equation}\label{C3}
    p^B_i = \sum_j p_{ij} = \sum_j p_{i|j}\, p^A_j \ .
\end{equation}
The above formula defines a classical channel ${\bf p}^B = T {\bf p}^A$, where the stochastic matrix $T$ is defined by $T_{ij} = p_{i|j}$. It is therefore clear that notions of a  classical channel and a classical conditional probability coincide

This correspondence suggests that one may introduce the concept of quantum conditional probability using well defined notion of a quantum channel, that is, completely positive trace preserving map (CPTP). Due to the Choi-Jamio{\l}kowski isomorphism \cite{CJ} the space of linear maps $\Lambda_{B|A} : \mathfrak{T}(\mathcal{H}_A) \longrightarrow \mathfrak{T}(\mathcal{H}_B)$ is isomorphic to $\mathfrak{B}(\mathcal{H}_A \ot \mathcal{H}_B)$: if $|i\>_A$ denote an orthonormal (computational) basis in $\mathcal{H}_A$ and $|\widetilde{\psi}^+_{AA}\> = \sum_i |ii\>_A$ stands for (unnormalized) maximally entangled vector in $\mathcal{H}_A \ot \mathcal{H}_A$, then one introduces
\begin{equation}\label{CJ}
\pi_{B|A} = (\oper_A \ot \Lambda_{B|A})\widetilde{P}^+_{AA} \ ,
\end{equation}
where $\widetilde{P}^+_{AA} = |\widetilde{\psi}^+_{AA}\>\<\widetilde{\psi}^+_{AA}|$. Suppose that $\Lambda_{B|A}$ is CPTP. Complete positivity implies that $\pi_{B|A} \geq 0$. Moreover, one has
\begin{eqnarray*}\label{}
 &&   {\rm Tr}_B\, \pi_{B|A} = {\rm Tr}_B \sum_{i,j} |i\>_A\<j| \ot \Lambda_{B|A}(|i\>_A\<j|) \\
 && = \sum_{i,j} |i\>_A\<j| \, {\rm Tr}[\Lambda_{B|A}(|i\>_A\<j|) ] \\
 && = \sum_{i,j} |i\>_A\<j| \, {\rm Tr}(|i\>_A\<j|) = \sum_i |i\>_A\<i| = \mathbb{I}_A\ ,
\end{eqnarray*}
where we have used ${\rm Tr}[\Lambda_{B|A}(\rho)] = {\rm Tr}\rho$. Following   \cite{Manolo,Leifer} let us introduce
\begin{definition} {\em We call $\pi_{B|A} \in \mathfrak{T}(\mathcal{H}_A\ot\mathcal{H}_B)$ a {\em quantum conditional probability} iff
\begin{equation}\label{Q1}
\pi_{B|A} \geq 0 \ ; \ \ {\rm Tr}_B\, \pi_{B|A} = \mathbb{I}_A\ .
\end{equation}
}
\end{definition}

\noindent In \cite{Leifer} $\pi_{B|A}$ is called {\em conditional quantum state}. Note, that (\ref{Q1}) is a quantum analog of (\ref{C1}).
It is therefore clear that formula (\ref{CJ}) establishes isomorphism between quantum channels and quantum conditional probability.

\begin{remark} It is often said that formula (\ref{CJ}) establishes isomorphism between quantum channels and quantum states. It is of course not true. Note that
\begin{equation}\label{CJ1}
\rho_{AB} = \frac{1}{d_A}\, (\oper_A \ot \Lambda_{B|A})\widetilde{P}^+_{AA} \ ,
\end{equation}
defines a legitimate state $\rho_{AB}$ in $\mathcal{H}_{AB}$. However, the inverse map
\begin{equation}\label{X}
    \Lambda_{B|A}(\rho) = d_A\, {\rm Tr}_A [ \rho_{AB} \cdot (\rho^T \ot \mathbb{I}_B)] \ ,
\end{equation}
is not trace preserving unless ${\rm Tr}_B\, \rho_{AB} = \mathbb{I}_A/d_A$, i.e. its marginal state $\rho_A$ is maximally mixed.
In this case $\pi_{B|A} = d_A\rho_{AB}$ defines quantum conditional probability.
\end{remark}
Note, that formula (\ref{X}) rewritten in terms of $\pi_{B|A}$
\begin{equation}\label{Q3}
    \Lambda_{B|A}(\rho) =  {\rm Tr}_A [ \pi_{B|A} \cdot (\rho^T \ot \mathbb{I}_B)] \ ,
\end{equation}
defines a quantum analog of (\ref{C3}). The transposition $\rho^T$ is performed with respect to the computational basis $|i\>_A$ in $\mathcal{H}_A$. One has
\begin{equation*}\label{}
    {\rm Tr}_A [ \pi_{B|A} \cdot (\rho \ot \mathbb{I}_B)] = {\rm Tr}_A [ (\rho^{\frac 12} \ot \mathbb{I}_B)\, \pi_{B|A}\,(\rho^{\frac 12} \ot \mathbb{I}_B)]\ ,
\end{equation*}
and hence a natural way to generalize (\ref{C2}) is to define a compound state $\rho_{AB}$ by the following symmetric prescription
\begin{equation}\label{}
    \rho_{AB} = (\rho^{\frac 12}_A \ot \mathbb{I}_B)\, \pi_{B|A}\,(\rho^{\frac 12}_A \ot \mathbb{I}_B)\ ,
\end{equation}
where $\rho_A$ is a state in $\mathcal{H}_A$. Note, that by construction ${\rm Tr}_B \, \rho_{AB}$ recovers $\rho_A$. Moreover
\begin{equation}\label{}
   \rho_B = {\rm Tr}_A \, \rho_{AB} = \Lambda_{B|A}(\rho_A^T)\ .
\end{equation}

\section{Quantum-correlation breaking channels}

Let us recall that an entanglement breaking channel \cite{EB} has the following well known Holevo representation
\begin{equation}\label{}
    \Lambda_{B|A}(\rho) = \sum_i {\rm Tr}(\rho F_i) \, R_i \ ,
\end{equation}
where $F_i$ stands for POVM in $\mathcal{H}_A$, and $R_i \geq 0$ together with ${\rm Tr}R_i = 1$. The corresponding conditional state
\begin{equation}\label{}
    \pi_{B|A} = \sum_i F^T_i \ot R_i \ ,
\end{equation}
is a separable positive operator in $\mathcal{H}_{AB}$. One has
\begin{equation}\label{}
    {\rm Tr}_B \, \pi_{B|A} = \sum_i F^T_i = \mathbb{I}_A\ ,
\end{equation}
by the very property of POVM. Now, a quantum conditional probability operator $\pi_{B|A}$ is quantum-classical (QC) [or more precisely $Q_AC_B$]  if $R_i$ mutually commute and hence there exists an orthonormal basis $|f_i\>$ in $\mathcal{H}_B$ such that
\begin{equation}\label{}
    R_i = \sum_j p_{j|i} |f_j\>\<f_j| \ ,
\end{equation}
where $p_{j|i}$ defines a  classical conditional probability. Indeed, $R_i$ is trivially positive and ${\rm Tr}\, R_i=1$ due to $\sum_jp_{j|i}=1$. Therefore, one has the following formula for QC conditional state
\begin{equation}\label{}
    \pi_{B|A} = \sum_{i,j} p_{j|i} F^T_i \ot |f_j\>\<f_j| \ ,
\end{equation}
and for the corresponding channel (quantum-to-classical measurement map \cite{Piani})
\begin{equation}\label{QC-channel}
    \Lambda_{B|A}(\rho) = \sum_i \, p_{j|i}\, {\rm Tr}(\rho F_i) \, |f_j\>\<f_j| \ .
\end{equation}
Hence the output state reads as follows
\begin{equation}\label{}
    \Lambda_{B|A}(\rho) = \sum_j p_j |f_j\>\<f_j| \ ,
\end{equation}
where $p_j = \sum_i p_{j|i} q_i$, with $q_i = {\rm Tr}(F_i \rho)$.

\begin{remark}
It shows that conditional probability $p_{i|j}$ appears already on the level of QC channels.
\end{remark}

A conditional state $\pi_{B|A}$ is called CQ [or more precisely $C_AQ_B$] iff $F_i$ mutually commute, that is, there exists an orthonormal basis $|e_i\>$ in $\mathcal{H}_A$ such that
\begin{equation}\label{}
    F_i = \sum_k q_{i|k} |e_k\>\<e_k| \ ,
\end{equation}
where $q_{i|k}$ stands for conditional probability. Hence, one has for CQ conditional state
\begin{equation}\label{}
    \pi_{B|A} = \sum_{i,j} \, q_{i|j}\, |e_j^*\>\<e_j^*| \ot R_i \ ,
\end{equation}
and for the corresponding channel
\begin{equation}\label{}
    \Lambda_{B|A}(\rho) = \sum_{i,j} \, q_{j|i}\, \<e_i|\rho|e_i\> \, R_j \ .
\end{equation}

Finally, $\pi_{B|A}$ is classical-classical (CC) if it is QC and CQ and hence
\begin{eqnarray}\label{}
    \pi_{B|A} &=& \sum_{i,j,k} \, p_{k|i}\, q_{i|j}\, |e_j^*\>\<e_j^*| \ot  |f_k\>\<f_k| \nonumber \\
    &=& \sum_{j,k} \, \pi_{k|j}\,  |e_j^*\>\<e_j^*| \ot  |f_k\>\<f_k|\ ,
\end{eqnarray}
where we have used
\begin{equation*}\label{}
    \pi_{k|j} = \sum_i \, p_{k|i}\, q_{i|j}\ .
\end{equation*}
It is clear that $\pi_{k|j}$ defines a legitimate conditional probability. The corresponding CC channel reads as follows
\begin{equation}\label{}
    \Lambda_{B|A}(\rho) = \sum_{j,k} \pi_{k|j} \<e_j|\rho|e_j\> \, |f_k\>\<f_k| \ .
\end{equation}

\begin{remark} Let $\Lambda : \mathfrak{T}(\mathcal{H}_A) \rightarrow \mathfrak{T}(\mathcal{H}_B)$ be a linear map (not necessarily completely positive) and let
\begin{equation}\label{CJ1}
 \Lambda \ \longrightarrow\    \sum_{i,j} |i\>_A\<j| \ot \Lambda(|i\>_A\<j|) \ ,
\end{equation}
denote the standard Choi-Jamio{l}kowski isomorphism.
Define a dual map $\Lambda^\# : \mathfrak{B}(\mathcal{H}_B) \rightarrow \mathfrak{B}(\mathcal{H}_A)$ via
\[  {\rm Tr} [ \Lambda^\#(a) \cdot \rho ] =  {\rm Tr} [  a \cdot \Lambda(\rho) ]\ . \]
For the dual map it is convenient to use the above isomorphism in the following ``dual" convention
\begin{equation*}\label{CJ2}
 \Lambda^\# \ \longrightarrow \   \sum_{k,l} \Lambda^\#(|k\>_B\<l|) \ot |k\>_B\<l| \ ,
\end{equation*}
since both (\ref{CJ1}) and (\ref{CJ2}) give rise to bi-partite operators in $\mathfrak{B}(\mathcal{H}_A \ot \mathcal{H}_B)$. Note, that
$\sum_{k,l} |k\>_B\<l| \ot \Lambda^\#(|k\>_B\<l|) \in  \mathfrak{S}(\mathcal{H}_B \ot \mathcal{H}_A)$.
\end{remark}

Let $\Lambda_{B|A}$ be a quantum channel. Its dual is not trace preserving unless $\Lambda_{B|A}$ is unital, i.e. $\Lambda_{B|A}(\mathbb{I}_A) = \mathbb{I}_B$. Suppose that $\Lambda_{B|A}(\mathbb{I}_A) = V > 0$ and define
\begin{equation}\label{}
    \widetilde{\Lambda}_{B|A}(\rho) = V^{-\frac 12} \Lambda_{B|A}(\rho) V^{-\frac 12}\ .
\end{equation}
Clearly, $\widetilde{\Lambda}_{B|A}$ is unital and hence its dual defines a quantum channel. Therefore, with each channel ${\Lambda}_{B|A}$, satisfying $\Lambda_{B|A}(\mathbb{I}_A) >0$, we may associate a channel $ \Lambda_{A|B} = \widetilde{\Lambda}_{B|A}^\#$ by
\begin{equation}\label{}
    \Lambda_{A|B}(\sigma) = {\Lambda}_{B|A}^\#(V^{-\frac 12} \sigma V^{-\frac 12})\ .
\end{equation}
Interestingly $\,\Lambda_{A|B}(V) = \mathbb{I}_A\,$. One has the following

\begin{thm}\label{T1}  ${\Lambda}_{B|A}$ is $Q_AC_B$ iff $\Lambda_{A|B}$ is $C_BQ_A$.
\end{thm}

\noindent {\bf Proof}: consider a $Q_AC_B$ channel defined in (\ref{QC-channel}). One has
\begin{equation}\label{}
    V = \Lambda_{B|A}(\mathbb{I}_A) = \sum_{i,j} p_{j|i}\, x_i |f_j\>\<f_j|\ ,
\end{equation}
with $x_i = {\rm Tr}\, F_i$, and assume that $p_j= \sum_i p_{j|i} x_i > 0$ for all $j=1,\ldots,d_B$. One obtains the following channel
\begin{equation}\label{}
    \Lambda_{A|B}(\sigma) = \sum_{i,j} \frac{p_{j|i}}{p_j} \, \<f_j|\rho|f_j\> \, F_i \ ,
\end{equation}
where $\sigma \in \mathfrak{S}(\mathcal{H}_B)$. Now, the classical Bayes theorem implies
\begin{equation*}\label{}
    p_{i|j} p_j = p_{j|i} x_i \ ,
\end{equation*}
and hence $\Lambda_{A|B}$ has a correct $C_BQ_A$ structure
\begin{equation}\label{}
    \Lambda_{A|B}(\sigma) = \sum_{i,j} {p_{i|j}} \, \<f_j|\rho|f_j\> \, R_i \ ,
\end{equation}
where $R_i = F_i/{x_i}  \,$ is a density operator in $\mathcal{H}_A$.  \hfill $\Box$

It is clear that any channel $\Lambda_{A|B} : \mathfrak{T}(\mathcal{H}_B) \longrightarrow \mathfrak{T}(\mathcal{H}_A)$ gives rise to a quantum conditional probability
\begin{equation}\label{}
    \pi_{A|B} = (\Lambda_{A|B} \ot \oper_B) \widetilde{P}^+_{BB}\ ,
\end{equation}
and the inverse map reads
\begin{equation}\label{}
    \Lambda_{A|B}(\sigma) = {\rm Tr}_B[ \pi_{A|B}^{T_B} (\mathbb{I}_A \ot \sigma)]\ .
\end{equation}

It should be clear that a concept of quantum conditional probability leads immediately to an analog of the Bayes theorem: let $\rho_{AB}$ be a bi-partite state in $\mathcal{H}_{AB}$ with reduced (marginal) states
\[    \rho_A = {\rm Tr}_B\, \rho_{AB}\ , \ \ \  \rho_B = {\rm Tr}_A\, \rho_{AB}\ , \]
satisfying $\rho_A >0$ and $\rho_B > 0$, that is, both reduced density operators are faithful. Let us introduce two quantum conditional probabilities
\begin{equation}\label{}
    \pi_{B|A} = (\rho_A^{- \frac 12} \ot \mathbb{I}_B)\, \rho_{AB}\, (\rho_A^{- \frac 12} \ot \mathbb{I}_B)\ ,
\end{equation}
and
\begin{equation}\label{}
    \pi_{A|B}  = (\mathbb{I}_A \ot \rho^{-\frac 12}_B)\, \rho_{AB}\,(\mathbb{I}_A \ot\rho^{-\frac 12}_B)\ .
\end{equation}
Therefore, by construction, one may formulate the quantum analog of Bayes theorem
\begin{eqnarray*}
      (\rho^{\frac 12}_A \ot \mathbb{I}_B)\, \pi_{B|A}\,(\rho^{\frac 12}_A \ot \mathbb{I}_B) = (\mathbb{I}_A \ot \rho^{\frac 12}_B)\, \pi_{A|B}\,(\mathbb{I}_A \ot \rho^{\frac 12}_B)\ .
\end{eqnarray*}
In particular
\begin{equation}\label{pi-pi}
\pi_{A|B} = (\rho^{\frac 12}_A \ot \rho^{-\frac 12}_B)\, \pi_{B|A}\,(\rho^{\frac 12}_A \ot \rho^{-\frac 12}_B)\ ,
\end{equation}
and if $\Lambda_{B|A}$ and $\Lambda_{A|B}$ denotes the corresponding quantum channels one easily finds the following relation
\begin{equation}\label{}
    \Lambda_{A|B}(\sigma) = \rho_A^{\frac 12} \Big[ \Lambda_{B|A}^\#\left(\rho_B^{-\frac 12} \sigma^T \rho_B^{-\frac 12} \right) \Big]^T      \rho_A^{\frac 12}\ ,
\end{equation}
where $\sigma \in \mathfrak{S}(\mathcal{H}_B)$. Note that
\begin{equation*}\label{}
    \Lambda_{B|A}(\rho_A) = \rho_B^T\ , \ \ \  \Lambda_{A|B}(\rho_B^T) = \rho_A\ .
\end{equation*}
In analogy to Theorem \ref{T1} one proves

\begin{thm} For any faithful $\rho_A$ and $\rho_B$ a channel ${\Lambda}_{B|A}$ is $Q_AC_B$ iff $\Lambda_{A|B}$ is $C_BQ_A$.
Equivalently, $\pi_{B|A}$ is  $Q_AC_B$ iff $\pi_{A|B}$ is $C_BQ_A$.
\end{thm}
Hence
\begin{cor}
For any faithful $\rho_A$ and $\rho_B$ a channel ${\Lambda}_{B|A}$ is $CC$ iff $\Lambda_{A|B}$ is $CC$.
Equivalently, $\pi_{B|A}$ is  $CC$ iff $\pi_{A|B}$ is $CC$.
\end{cor}

Now, we reformulate the main result of \cite{KHH} in terms of quantum conditional probability.

\begin{thm}
Let $\Lambda_{B|A}$ be a quantum channel. The corresponding quantum conditional probability
$\pi_{B|A} = (\oper_A \ot \Lambda_{B|A})\widetilde{P}^+_{AA}$ is $Q_AC_B$  if and only if $(\oper_C \ot \Lambda_{B|A})\rho_{CA}$
is a $Q_CC_B$ state in $\mathcal{H}_{CB}$ for any bipartite state $\rho_{CA}$ in $\mathcal{H}_{CA}$.
\end{thm}

\noindent {\bf Proof}: suppose that $\pi_{B|A}$ is $Q_AC_B$, that is,
\begin{equation}\label{}
    \Lambda_{B|A}(\rho) = \sum_{i,j} p_{j|i} \, {\rm Tr}(\rho F_i) \, |f_j\>\<f_j| \ .
\end{equation}
Let
\begin{equation}\label{}
    \rho_{CA} = \sum_{i,j} \rho_{ij} \ot |i\>_A\<j| \ ,
\end{equation}
with $\rho_{ij} \in \mathfrak{T}(\mathcal{H}_C)$.
One has
\begin{eqnarray}
 && (\oper_C \ot \Lambda_{B|A})\rho_{CA}  = \sum_{i,j} \rho_{ij} \ot \Lambda_{B|A}(|i\>\<j|) \nonumber \\
 && =  \sum_{i,j}\sum_{k,l}\, p_{k|l}\, \rho_{ij} \ot {\rm Tr}(F_l |i\>_A\<j|)|f_k\>\<f_k|  \nonumber  \\
 && = \sum_{k,l} p_{k|l}\, \sigma_l \ot |f_k\>\<f_k|\ ,
\end{eqnarray}
where
\begin{equation}\label{sigma-l}
    \sigma_l = \sum_{i,j}\,  \rho_{ij} \<j|F_l |i\>\ ,
\end{equation}
defines a set of positive operators in  $\mathcal{H}_C$. Let $p_l = {\rm Tr}\, \sigma_l$ and let $p_{kl} = p_{k|l}p_l$. One has
\begin{equation}\label{}
    (\oper_C \ot \Lambda_{B|A})\rho_{CA} = \sum_{k,l} p_{kl}\, \rho_l \ot |f_k\>\<f_k|\ ,
\end{equation}
where $\rho_l = \sigma_l/p_l$ are density operators in $\mathcal{H}_C$. \hfill $\Box$

\begin{remark}
The positivity of $\sigma_l$ is not very clear from (\ref{sigma-l}).
Note, that
\begin{equation*}\label{}
    (\mathbb{I}_C \ot F_l^{\frac 12})\, \rho_{CA}\, (\mathbb{I}_C \ot F_l^{\frac 12})  = \sum_{i,j} \rho_{ij} \ot F_l^{\frac 12}|i\>\<j|F_l^{\frac 12}   \ ,
\end{equation*}
is evidently positive. Hence
\begin{equation*}\label{}
    {\rm Tr}_A [(\mathbb{I}_C \ot F_l^{\frac 12})\, \rho_{CA}\, (\mathbb{I}_C \ot F_l^{\frac 12})]  = \sum_{i,j} \rho_{ij} \<j|F_l |i\> \ ,
\end{equation*}
is positive as well. Moreover $p_l = {\rm Tr} \sigma_l \leq 1$.
\end{remark}

Let $\Lambda_{B|A}$ be CC. As was shown in \cite{KHH} it is not true that $(\oper_C \ot \Lambda_{B|A})\rho_{CA}$ is CC for arbitrary  $\rho_{CA}$. Following \cite{KHH} denote by $CC(\Lambda_{B|A})$ a set of states in $\mathcal{H}_{CA}$ such that $(\oper_C \ot \Lambda_{B|A})\rho_{CA}$ is $C_CC_B$.

\begin{thm}
Let $\Lambda_{B|A}$ be CC
\begin{equation}\label{CC1}
    \Lambda_{B|A}(\rho) = \sum_{i,j} \pi_{j|i} \, \<e_i|\rho|e_i\> \, |f_j\>\<f_j| \ .
\end{equation}
A state $(\oper_C \ot \Lambda_{B|A})\rho_{CA}$ in $\mathcal{H}_{CB}$ is CC iff the diagonal blocks $\rho_{ii}$ of the following representation
\begin{equation}\label{}
    \rho_{CA} = \sum_{i,j} \rho_{ij} \ot |e_i\>\<e_j| \ ,
\end{equation}
mutually commute.
\end{thm}

\noindent {\bf Proof}: one has
\begin{eqnarray}
 && (\oper_C \ot \Lambda_{B|A})\rho_{CA}  = \sum_{i,j} \rho_{ij} \ot \Lambda_{B|A}(|e_i\>\<e_j|) \nonumber \\
 && =  \sum_{i,j}\sum_{k,l}\, \pi_{k|l}\, \rho_{ij} \ot \<e_l|e_i\>\<e_j|e_l\> |f_k\>\<f_k|  \nonumber  \\
 && = \sum_{k,l} \pi_{k|l}\, \rho_{ll} \ot |f_k\>\<f_k|\ ,
\end{eqnarray}
which is CC iff $[\rho_{ll},\rho_{kk}]=0$. \hfill $\Box$

\vspace{.1cm}

Let us observe that the above Theorem immediately reproduces results from \cite{KHH}. Indeed, let $|\psi_{CA}\> \in \mathcal{H}_C \ot \mathcal{H}_A$ and $|\psi_{CA}\> = \sum_i c_i |\widetilde{e}_i\> \ot |e_i\>$ be its Schmidt decomposition. One has $\rho_{CA} = |\psi_{CA}\>\<\psi_{CA}| = \sum_{i,j} c_i \overline{c}_j |\widetilde{e}_i\>\<\widetilde{e}_j| \ot |e_i\>\<e_j|$ and hence $\rho_{ii} = |c_i|^2 |\widetilde{e}_i\>\<\widetilde{e}_i|$ mutually commute. Similarly, let us consider a CQ state in $\mathcal{H}_C \ot \mathcal{H}_A$, i.e.
$\rho_{CA} = \sum_k p_k |\widetilde{e}_k\>\<\widetilde{e}_k| \ot \sigma_k$. One has
$$ \rho_{CA} = \sum_{i,j} \sum_k p_k |\widetilde{e}_k\>\<\widetilde{e}_k| \ot \< {e}_i|\sigma_k|e_j\>  |e_i\>\<e_j|\ , $$
and hence $\rho_{ii} =  \sum_k p_k \< {e}_i|\sigma_k|e_i\>\, |\widetilde{e}_k\>\<\widetilde{e}_k|$ mutually commute.

\vspace{.1cm}

It is therefore clear that $\rho_{CA}$ has the following form
\begin{equation}\label{}
    \rho_{CA} =  \sigma'_{CA}  + \sigma''_{CA} \ ,
\end{equation}
where
\begin{equation}\label{}
    \sigma'_{CA} = \sum_{i,j} \tau_{ji}  \, |c_i\>\<c_i| \ot |e_j\>\<e_j|\ ,
\end{equation}
is a CC state ($|c_i\>$  stands for the orthonormal basis in $\mathcal{H}_C$), and
\begin{equation}\label{}
    \sigma''_{CA} = \sum_{i\neq j} \rho_{ij} \ot |e_i\>\<e_j|\ .
\end{equation}
The off-diagonal part $\sigma''_{CA}$ is arbitrary provided $\rho_{CA} \geq 0$. The channel (\ref{CC1}) acting upon $\rho_{CA}$ kills $\sigma''_{CA}$ and transforms CC part $\sigma'_{CA}$ into
\begin{equation*}\label{}
   (\oper_C\ot \Lambda_{B|A})\sigma''_{CA} = \sum_{j,k} \, \kappa_{jk}\,  |c_k\>\<c_k| \ot |f_j\>\<f_j| \ ,
\end{equation*}
where $\kappa_{jk} =  \sum_{i}  \pi_{j|i}\, \tau_{ik}$ is a legitimate joint probability distribution.
Let $P_i = |e_i\>\<e_i|$. Define a CPTP projector
\begin{equation*}\label{}
    \mathcal{P}_A(\rho_A) = \sum_i \, P_i \rho_{A}  P_i\ .
\end{equation*}

\begin{cor}
A state $\rho_{CA} \in CC(\Lambda_{B|A})$ iff $[\oper_C \ot \mathcal{P}_A](\rho_{CA})$ is $C_CC_A$.
\end{cor}
This result may be reformulated in terms of partial decoherence. Let
\begin{equation*}\label{}
    L = \gamma (\oper_A - \mathcal{P}_A)\ ,
\end{equation*}
denote a genuine generator of Markovian semigroup in $\mathfrak{T}(\mathcal{H}_A)$. One finds for the corresponding dynamical map $\Phi^A_t = e^{tL}$:
\begin{equation*}\label{}
  \rho_t =   \Phi^A_t(\rho) = e^{-\gamma t} \rho + (1-e^{-\gamma t}) \mathcal{P}_A(\rho)\ ,
\end{equation*}
and hence the asymptotic state
\begin{equation*}\label{}
    \rho_\infty = \mathcal{P}_A(\rho)\ ,
\end{equation*}
is perfectly decohered with respect to $|e_i\>$ basis in $\mathcal{H}_A$. Consider now the following partial decoherence
in $\mathcal{H}_{CA}$
\begin{equation*}\label{}
    \Phi^{CA}_t = \oper_C \ot \Phi^A_t\ .
\end{equation*}
It is clear that $\rho_{CA} \in CC(\Lambda_{B|A})$ iff
the partially decohered asymptotic state $\Phi^{CA}_\infty(\rho_{CA})$ is $C_CC_A$.

\section{Quantum Perron-Frobenius theorem and broadcastability of quantum states}


Authors of \cite{KHH} found an interesting connection between broadcastability of quantum states by QC channels and celebrated Perron-Frobenius theory. In this section we are going to explore this connection further. It turns out that considering arbitrary channels (not necessarily  QC ones) we need to consider the quantum analog of Perron-Frobenius theory \cite{P-F1,P-F2} (recently Perron-Frobenius theory was used in the analysis of spectra of random quantum channels \cite{PF}).

Let $\mathcal{H}_A = \mathcal{H}_B = \mathcal{H}$. A state $\rho_*$ in $\mathcal{H}$ is broadcastable by the channel $\Lambda$ if the corresponding compound state
\begin{equation}\label{}
    \rho_{AB} = (\rho_*^{\frac 12} \ot \mathbb{I}_B)\, \pi_{B|A}\, (\rho_*^{\frac 12} \ot \mathbb{I}_B)\ ,
\end{equation}
provides a broadcast for $\rho_*$, that is, $\rho_A=\rho_B=\rho_*$.  It is clear that by construction $\rho_A = \rho_*$ and $\rho_B = \Lambda_{B|A}(\rho_*^T)$. It should be stressed that this is a standard definition of broadcastability {\em via} a bipartite broadcast state $\rho_{AB}$. We provide only special construction of $\rho_{AB}$ using a quantum conditional probability operator $\pi_{B|A}$.

\begin{cor} A state $\rho_*$ is broadcastable by the channel $\Lambda$ iff
\begin{equation}\label{}
    \Lambda^\tau(\rho_*) =\rho_*\ ,
\end{equation}
that is, $\rho_*$ is a fixed point of the positive trace preserving map $ \Lambda^\tau =  \Lambda \circ T$, where again $T$ denotes transposition with respect to the computational basis in $\mathcal{H}_A$.
\end{cor}

Let  $\Phi : \mathfrak{T}(\mathcal{H}) \rightarrow \mathfrak{T}(\mathcal{H})$ be a positive trace-preserving map (not necessarily completely positive). One proves \cite{P-F1,P-F2} that  $\Phi$ possesses an eigenvalue $\lambda_*=1$ and the corresponding (in general not unique) eigenvector $\rho_*$ (after suitable normalization) defines a legitimate density operator in $\mathcal{H}$. Moreover, the remaining (in general complex) eigenvalues $\lambda_\alpha$ satisfy $|\lambda_\alpha| \leq |\lambda_*|=1$. To guarantee the uniqueness of $\rho_*$ one needs extra conditions upon $\Phi$. The quantum analog of irreducibility reads as follows: $\Phi$ is irreducible iff $(\oper + \Lambda)^{n-1}(\rho) > 0$ for all $\rho \geq 0$ and $n = {\rm dim}\, \mathcal{H}$. If $\Phi$ is irreducible then $\rho_*$ is unique. Finally, let us call $\Phi$ {\em primitive} (or {\em regular}) iff for some integer $k$ one has $\Phi^k(A) > 0$ for all $A \geq 0$. If $\Phi$ is primitive then clearly the Perron-Frobenius vector $\rho_*$ is unique and moreover the remaining eigenvalues $\lambda_\alpha$ satisfy $|\lambda_\alpha| < |\lambda_*|=1$.

\begin{cor} \label{cor4} For an arbitrary  quantum channel $\Lambda$ there exists a state $\rho_*$ (Perron-Frobenius eigenvector of $\Lambda^\tau$)  which is broadcastable by $\Lambda$.
\end{cor}

Let us consider the spectral problem for $\Lambda^\tau$ and its dual ${\Lambda^\tau}^{\#}$:
\begin{equation}\label{}
    \Lambda^\tau (X_\alpha) = \lambda_\alpha X_\alpha\ , \ \ \  {\Lambda^\tau}^\# (Y_\alpha) = \lambda^*_\alpha Y_\alpha\ ,
\end{equation}
with $\alpha=0,1,\ldots,d^2-1$ ($d={\rm dim}\, \mathcal{H}$).  The so called damping basis \cite{Briegel} satisfies ${\rm Tr}(X_\alpha Y_\beta^\dagger)=\delta_{\alpha\beta}$. One has $\lambda_0=1$, $X_0=\rho_*$ and $Y_0 = \mathbb{I}$. Moreover, ${\rm Tr}\, X_\alpha = 0$ for all $\alpha >0$. The action of $\Lambda^\tau$ may be, therefore, represented as follows
\begin{eqnarray}\label{}
    \Lambda^\tau(\rho) &=& \sum_{\alpha=0}^{d^2-1} \lambda_\alpha X_\alpha {\rm Tr}(Y_\alpha^\dagger \rho)
    =  \rho_* {\rm Tr}\, \rho  + \xi(\rho) \ ,
\end{eqnarray}
where the traceless operator $\xi(\rho)$ reads
\begin{eqnarray*}\label{}
    \xi = \sum_{\alpha>0} \lambda_\alpha X_\alpha {\rm Tr}(Y_\alpha^\dagger \rho) \ .
\end{eqnarray*}
Equivalently, the original channel $\Lambda$ acts as follows
\begin{eqnarray*}\label{}
    \Lambda(\rho) = \sum_{\alpha=0}^{d^2-1} \lambda_\alpha X_\alpha {\rm Tr}(Y_\alpha^\dagger \rho^T)
    =  \rho_* {\rm Tr}\, \rho  + \xi^\tau(\rho) \ ,
\end{eqnarray*}
with the traceless $ \xi^\tau(\rho)  = \sum_{\alpha>0}^{d^2-1} \lambda_\alpha X_\alpha {\rm Tr}(Y_\alpha^\dagger \rho^T)$.

The corresponding quantum conditional probability is given by the following formula
\begin{eqnarray}\label{}
    \pi_{B|A} &=& \sum_{i,j} |i\>_A\<j| \ot \Lambda(|i\>_A\<j|) = \sum_{\alpha=0}^{d^2-1} \lambda_\alpha Y^\dagger_\alpha \ot X_\alpha\nonumber \\ &=& \mathbb{I}_A \ot \rho_* +  \sum_{\alpha >0 } \lambda_\alpha Y^\dagger_\alpha \ot X_\alpha\ .
\end{eqnarray}
One clearly sees that ${\rm Tr}_B \pi_{B|A} = \mathbb{I}_A$ due to the normalization ${\rm Tr}\rho_* = 1$ and ${\rm Tr}\, X_\alpha =0$ for $\alpha >0$. On the other hand one has ${\rm Tr}_A \pi_{B|A} = \sum_\alpha \lambda_\alpha {\rm Tr}(Y^\dagger_\alpha)\, X_\alpha = \Lambda(\mathbb{I}_A)$. Hence, if $\Lambda$ is unital, then  ${\rm Tr}_A \pi_{B|A} = \mathbb{I}_B$. The compound $\rho_{AB}$ state is given by
\begin{eqnarray}\label{CAN}
    \rho_{AB} = (\rho_*^\frac 12 \ot \mathbb{I}_B)  \pi_{B|A} (\rho_*^\frac 12 \ot \mathbb{I}_B)  = \rho_* \ot \rho_* + \zeta_{AB} \ ,
\end{eqnarray}
with
\begin{equation}\label{}
    \zeta_{AB} = \sum_{\alpha >0 } \lambda_\alpha \, \rho_*^\frac 12 \, Y^\dagger_\alpha \rho_*^\frac 12 \ot X_\alpha\ .
\end{equation}
Note, that ${\rm Tr}_A \zeta_{AB} = {\rm Tr}_B \zeta_{AB} = 0$.
The formula (\ref{CAN}) may be therefore considered as the canonical representation of the broadcast for $\rho_*$.

\begin{cor} A compound state $\rho_{AB}$ is a broadcast for $\rho_*$ iff
$$   \zeta_{AB} = \rho_{AB} - \rho_* \ot \rho_*\ , $$
satisfies ${\rm Tr}_A \zeta_{AB} = {\rm Tr}_B \zeta_{AB} = 0$.
\end{cor}

It should be stressed that Corollaries 3 and 5 follow directly
from the very definition of broadcastability. Authors of \cite{KHH} introduced a notion of spectrum broadcastability: $\widetilde{\rho}_*$ is spectrum broadcastable by $\Lambda$ if there exists unitary $U$ such that
\begin{equation}\label{}
    \rho_B = U \widetilde{\rho}_* U^\dagger\ ,
\end{equation}
that is, $\rho_B$ has the same spectrum as $\widetilde{\rho}_*$ (again $\rho_A=\widetilde{\rho}_*$).

Let $U$ be an arbitrary unitary operator in $\mathcal{H}$. Denote by $\Lambda_U^\tau$ the following positive and trace preserving map
\begin{equation}\label{}
    \Lambda^\tau_U(\rho) = U^* \Lambda(\rho^T) U^T \ .
\end{equation}
Let $\rho_*^U$ be the corresponding Perron-Frobenius eigenvector, that is,
\begin{equation}\label{}
    \Lambda^\tau_U(\rho^U_*) = \rho_*^U\ .
\end{equation}

\begin{pro} The state $\rho^U_*$ is spectrum broadcastable by $\Lambda$.
\end{pro}

\noindent {\bf Proof}: one has
\begin{eqnarray}\label{CAN-U}
    \rho^U_{AB} &=& ({\rho_*^U}^\frac 12 \ot \mathbb{I}_B)  \pi_{B|A} ({\rho_*^U} ^\frac 12 \ot \mathbb{I}_B) \nonumber \\
     &=& \rho_*^U \ot \rho_* + \zeta^U_{AB} \ ,
\end{eqnarray}
with
\begin{equation}\label{}
    \zeta_{AB}^U = \sum_{\alpha >0 } \lambda_\alpha \, {\rho_*^U}^\frac 12 \, Y^\dagger_\alpha {\rho_*^U}^\frac 12 \ot X_\alpha\ .
\end{equation}
This gives
$$   \rho_A = \rho_*^U \ , \ \ \ \rho_B = \Lambda^\tau(\rho_*^U) = U \rho_*^U U^\dagger \ , $$
which proves spectrum broadcastability. \hfill $\Box$

Note, that if $\Lambda^\tau$ is a primitive map, i.e. $\rho_*$ is unique and  $|\lambda_\alpha|<|\lambda_0|=1 $ for  $\alpha >0$, then there exists a limit
\begin{equation}\label{}
    \lim_{r\rightarrow \infty} \Lambda^r = \Lambda_\infty\ ,
\end{equation}
defined by $\Lambda_\infty(\rho) = \rho_* {\rm Tr}\rho$. In this case $\Lambda_\infty$ breaks all correlations present in an arbitrary state
$(\Lambda \ot \oper_B)\rho_{AB} = \rho_* \ot \rho_B$ with $\rho_B = {\rm Tr}\rho_{AB}$. Again this result holds for an arbitrary channel $\Lambda$ (not necessarily QC one \cite{KHH}).

\section{Conclusions}

We provided characterization of correlation breaking channels in terms of quantum conditional probability (or quantum conditional states). Using the quantum analog of Bayes theorem which relates $\pi_{B|A}$ and $\pi_{A|B}$ it was shown that any $Q_AC_B$ channel $\Lambda_{B|A}$ gives rise to the whole family on $C_BQ_A$ channels $\Lambda_{A|B}$. All these channels enjoy the following property: if $\rho_B = \Lambda_{B|A}(\rho_A)$, then $\rho_A = \Lambda_{A|B}(\rho_B)$. Interestingly, these channels corresponds to the Barnum-Knill recovery channels \cite{Barnum}, that is, $\Lambda_{A|B}$ recovers `quantum' $\rho_A$ out of the `classical' $\rho_B$.

Finally, using the quantum analog of celebrated Perron-Frobenius theorem we provided generalization of results of Korbicz et. al. \cite{KHH} from QC channels to arbitrary quantum channels. A quantum channel $\Lambda$ may be used to broadcast the Perron-Frobenius eigenvector of $\Lambda^\tau = \Lambda \circ T$. Moreover, there exists a whole family of states spectrum broadcastable by $\Lambda\,$: for each unitary operator $U$ the corresponding Perron-Frobenius eigenvector of $\Lambda^\tau_U$ is spectrum broadcastable by the original channel $\Lambda$.

The use of quantum conditional probability enables one to reproduce all results of \cite{KHH}. In particular Theorem 4 generalizes the characterization of the set of bi-partite states $\rho_{CA}$ such that $[\oper_C \ot \Lambda_{B|A}](\rho_{CA})$  is a CC state for a given CC channel $\Lambda_{B|A}$.

To summarize:  there exist an intriguing connections between quantum analogs of conditional probability, Bayes theorem and Perron-Frobenius theorem. We believe that these connections deserve further analysis.



\section*{Acknowledgements}

I thank Jarek Korbicz for discussions. This work was partially supported by the National Science Center project  
DEC-2011/03/B/ST2/00136.

\end{document}